\documentclass[runningheads]{llncs}
\usepackage{graphicx}
\usepackage{todonotes}
\usepackage{subfig}
\graphicspath{{figures/}}
\usepackage[english]{babel}
\usepackage{hyperref}

\usepackage[autostyle=true,english=american]{csquotes}
\usepackage{acro}
\DeclareAcronym{UI}{short = UI, long = user interface}
\DeclareAcronym{GUI}{short = GUI, long = graphical user interface}
\DeclareAcronym{TLX}{short = NASA TLX, long = NASA-Task Load Index}
\DeclareAcronym{RTLX}{short = NASA Raw-TLX, long =NASA Raw-Task Load Index}
\DeclareAcronym{ER}{short = ER, long = error rate}
\DeclareAcronym{TCT}{short = TCT, long = task completion time}
\DeclareAcronym{HCI}{short = HCI, long = Human-Computer Interaction}
\DeclareAcronym{UX}{short = UX, long = user experience}
\DeclareAcronym{RMSE}{short = RMSE, long = root mean squared error}
\DeclareAcronym{HMD}{short = HMD, long = Head-Mounted Display, long-plural-form = Head-Mounted Displays}
\DeclareAcronym{CNN}{short = CNN, long = Convolutional Neural Network}
\DeclareAcronym{FOV}{short = FoV, long = field of view}
\DeclareAcronym{HRC}{short = HRC, long = Human-Robot Collaboration}
\DeclareAcronym{HRI}{short = HRI, long = Human-Robot Interaction}
\DeclareAcronym{ANOVA}{short = ANOVA, long = analysis of variance}
\DeclareAcronym{RMANOVA}{short = RM-ANOVA, long = Repeated Measures Analysis of Variance}
\DeclareAcronym{JND}{short = JND, long =just-noticeable difference}
\DeclareAcronym{SUS}{short = SUS, long =system usability scale}
\DeclareAcronym{CSCW}{short = CSCW, long = computer-supported cooperative work}
\DeclareAcronym{CAD}{short = CAD, long = computer-aided design}
\DeclareAcronym{MR}{short = MR, long = Mixed Reality}
\DeclareAcronym{AR}{short = AR, long = Augmented Reality}
\DeclareAcronym{AV}{short = AV, long = Augmented Virtuality}
\DeclareAcronym{VR}{short = VR, long = Virtual Reality}
\DeclareAcronym{SAR}{short = SAR, long = Spatial Augmented Reality}
\DeclareAcronym{ADLs}{short = ADLs, long = Activities of Daily Living}
\DeclareAcronym{LED}{short = LED, long = Light-Emitting Diode}
\DeclareAcronym{DoF}{short = DoF, long = Degree-of-Freedom, long-plural-form = Degrees-of-Freedom}
\DeclareAcronym{HHC}{short = HHC, long = Human-Human Collaboration}
\DeclareAcronym{AI}{short = AI, long = Artifical Intelligence}
\DeclareAcronym{QUEAD}{short = QUEAD, long = Questionnaire for the Evaluation of Physical Assistive Devices}
\DeclareAcronym{TiA}{short = TiA, long = Trust in Automation Questionnaire}
\DeclareAcronym{TOR}{short = TOR, long = Take-Over-Request}
\DeclareAcronym{ADMC}{short = ADMC, long = Adaptive DoF Mapping Controls}
\DeclareAcronym{ROS}{short = ROS, long = Robot Operating System}
\DeclareAcronym{TCP}{short = TCP, long = Tool Center Point}
\DeclareAcronym{DnD}{short = D\&D, long = Design and Development}
\DeclareAcronym{XR}{short = XR, long= Extended Reality}

\begin{document}
\title{Exploring AI-enhanced Shared Control for an Assistive Robotic Arm\thanks{This research is supported by the \textit{German Federal Ministry of Education and Research} (BMBF, FKZ: \href{https://foerderportal.bund.de/foekat/jsp/SucheAction.do?actionMode=view&fkz=16SV8565}{16SV8565}).}}

\author{Max Pascher\inst{1,2}\orcidID{0000-0002-6847-0696}
\and \\Kirill Kronhardt\inst{1}\orcidID{0000-0002-0460-3787}
\and \\Jan Freienstein\inst{3}\orcidID{0009-0006-4287-8980}
\and \\Jens Gerken\inst{1}\orcidID{0000-0002-0634-3931}
}
\authorrunning{M. Pascher et al.}
\institute{TU Dortmund University, Dortmund, NW, Germany 
\and University of Duisburg-Essen, Essen, NW, Germany 
\and Westphalian University of Applied Sciences, Gelsenkirchen, NW, Germany}
\maketitle              % typeset the header of the contribution
\begin{abstract}
Assistive technologies and in particular assistive robotic arms have the potential to enable people with motor impairments to live a self-determined life. More and more of these systems have become available for end users in recent years, such as the \emph{Kinova Jaco} robotic arm. However, they mostly require complex manual control, which can overwhelm users. As a result, researchers have explored ways to let such robots act autonomously. However, at least for this specific group of users, such an approach has shown to be futile. Here, users want to stay in control to achieve a higher level of personal autonomy, to which an autonomous robot runs counter. In our research, we explore how \ac{AI} can be integrated into a shared control paradigm. In particular, we focus on the consequential requirements for the interface between human and robot and how we can keep humans in the loop while still significantly reducing the mental load and required motor skills.

\keywords{Assistive Technologies \and Human-Robot-Interaction Mixed Reality \and Shared Control \and Visual Cues.}
\end{abstract}
\section{Introduction and Motivation}
\label{sec:intro}
When controlling an assistive robotic arm, one major challenge from a human-robot interface perspective is the mapping between available input controls and resulting robot movements. Assistive robotic arms capable of performing arbitrary pick-and-place tasks in 3D-space require at least seven \acp{DoF}: x-, y- and z-translation, roll, pitch, yaw, and opening / closing the robot's fingers (\emph{cardinal} \acp{DoF}). Therefore, there is no direct mapping possible with most input devices. As a result, the user needs to constantly switch between modes, i.e., flip through several pre-defined mappings between input and output space. 
Research has shown that such mode switching requires a considerable amount of time and cognitive demand~\cite{Herlant.2016modeswitch,Chung.2013,Al-Halimi2017}.

Much research in assistive robotics is concerned with autonomous robotic functions~\cite{Canal.2016,lauretti2017learning,Gallenberger.2019,rakhimkul2019autonomous}. From that perspective, the mode switching dilemma may seem moot with robots gaining more and more autonomy through advanced \acf{AI} and thereby reducing the need for such direct control altogether. However, studies have shown that people with motor impairments may prefer manual control over autonomous execution, as they see the robot not so much as another agent but as a tool to regain self-determination (e.g.,~\cite{Kim.2012,Pollak.2020autonomystress}).  

To tackle this dilemma, different approaches along the continuum of shared control methods were proposed (e.g.,~\cite{Goldau.2021petra,Kronhardt.2022adaptOrPerish,Herlant.2016modeswitch,tsui2011want,ezeh2017,quere2020,jain2015,Pascher2023c}). The concept of shared control has great potential to design communication and control between humans and robots~\cite{Abbink2018}. 
For assistive robotic arms, however, the various approaches address the issue on different levels, sometimes reducing the involvement of the user to simply indicate an object to be picked~\cite{tsui2011want}. Other approaches directly adress the mode switching issue, with Herlant et al. suggesting time-optimal mode switching along the cardinal \acp{DoF}~\cite{Herlant.2016modeswitch} and -- based on Dijkstra's algorithm -- to predict when the robot should switch modes.

Recently, Goldau \& Frese proposed an approach integrating a \ac{CNN}, which interprets live camera data from the gripper to constantly describe the probabilistic distribution of intended \ac{DoF} robot motion and, accordingly, optimal mapping of \acp{DoF}~\cite{Goldau.2021petra}. In principle, the idea is that the user gets a suggestion not just for when and how to switch mode but going beyond cardinal \acp{DoF} suggestions, allowing more flexible and adaptive \ac{DoF} combinations.

In our work, we explored the feasibility of such a \ac{CNN}-based approach, identified empirical implications for shared control systems, and what kind of human-robot interaction design is feasible~\cite{Kronhardt.2022adaptOrPerish,Pascher.2024adaptix,Pascher2023c}. For this paper, we present the main challenges we identified and how our work has aimed to address these. In summary, these challenges are:

\begin{itemize}
    \item \textbf{\ac{AI} legibility:} Given an \ac{AI} which is able to automate mode switches, the user needs to understand the behavior and actions of the \ac{AI}. A system which changes input mappings without notification or explanation would be perceived as unpredictable.
    \item \textbf{\ac{AI} user control:} Given an \ac{AI} which is able to automate mode switches, the user must stay in the loop and have a final say in making the choices.
    
    \item \textbf{\ac{AI} intervention:} Given an \ac{AI} which is able to automate mode switches, the user needs to expect and prepare for \ac{AI} mistakes. Therefore, they need to interfere with the \ac{AI} and, at best, make it reconsider the mode switching.
\end{itemize}

\section{Engineering Context: Simulation Environment}
In our research, we found that -- given the complexity of robotic arms in combination with the limitations of current \ac{AI} technologies -- a testbed environment that allows the integration of different control mechanisms and user interface components facilitates engineering.

To that end, we developed \emph{AdaptiX}~\cite{Pascher.2024adaptix}, a transitional \acs{XR} framework for shared control applications in assistive robotics. 
\emph{AdaptiX} resembles a real-world scenario, where an assistive robotic arm (here a \emph{Kinova Jaco 2}) is used to facilitate pick-and-place tasks as we observed that they are part of many \ac{ADLs}. 
The system combines a physical robot implementation with a 3D simulation environment. This approach, reminiscent of simulations used in industrial contexts~\cite{Matsas.2015vrIndustry,Muller.2017vrIndustry,Tidoni.2017vrIndustry}, helps to address challenges associated with bulky, expensive, and complex assistive robotic arms.
Researchers are empowered to streamline their \ac{DnD} process, reducing complexity and enhancing efficiency. The system's integrated \ac{ROS} interface enables seamless connectivity to a physical robotic arm, supporting bidirectional interactions and data exchange through a \emph{DigitalTwin} and \emph{PhysicalTwin} approach.

\begin{figure}[htbp]
    \centering
    \includegraphics[width=\linewidth]{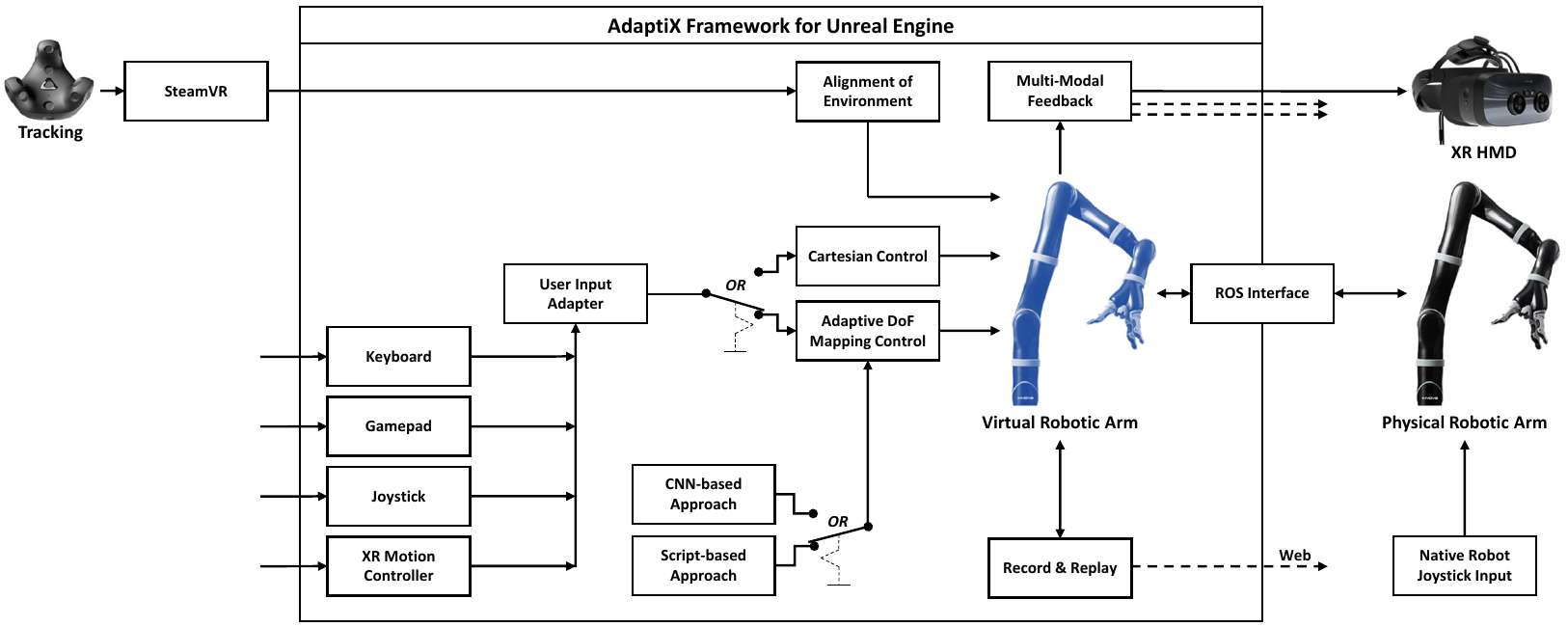}
    \caption{Overview of \emph{AdaptiX}' architecture, illustrating each component, their directional communication, and the crossover from and to the framework~\cite{Pascher.2024adaptix}.}% The user input is either used for \emph{Cartesian Control} or \ac{ADMC}. For \acs{ADMC}, either a \ac{CNN}-based or script-based rule engine can be selected.}
    \label{fig:framework-architecture}
\end{figure}

In addition to Cartesian robot control, the framework includes \ac{ADMC}, an initial shared control approach that employs \ac{AI}-generated suggestions, subject to user approval and control. \autoref{fig:framework-architecture} provides an overview of the framework's architecture.

\emph{AdaptiX} is built on the foundation of the game engine \emph{Unreal Engine 4.27}~\cite{unrealengine}. This game engine is renowned for its advanced real-time 3D photorealistic visuals and immersive capabilities, making it an ideal choice for our framework. Furthermore, it offers a rich set of assets that can be readily used for future expansions. \emph{Unreal Engine} is versatile and supports a variety of hardware configurations, allowing the framework to be deployed across different operating systems. It is compatible with a wide range of \ac{VR}, \ac{MR}, and \ac{AR} headsets, as well as gamepads and joysticks, making it suitable for development in both \emph{C++} and \emph{Blueprints}.

\begin{figure}[htbp]
\centering
\includegraphics[width=\linewidth]{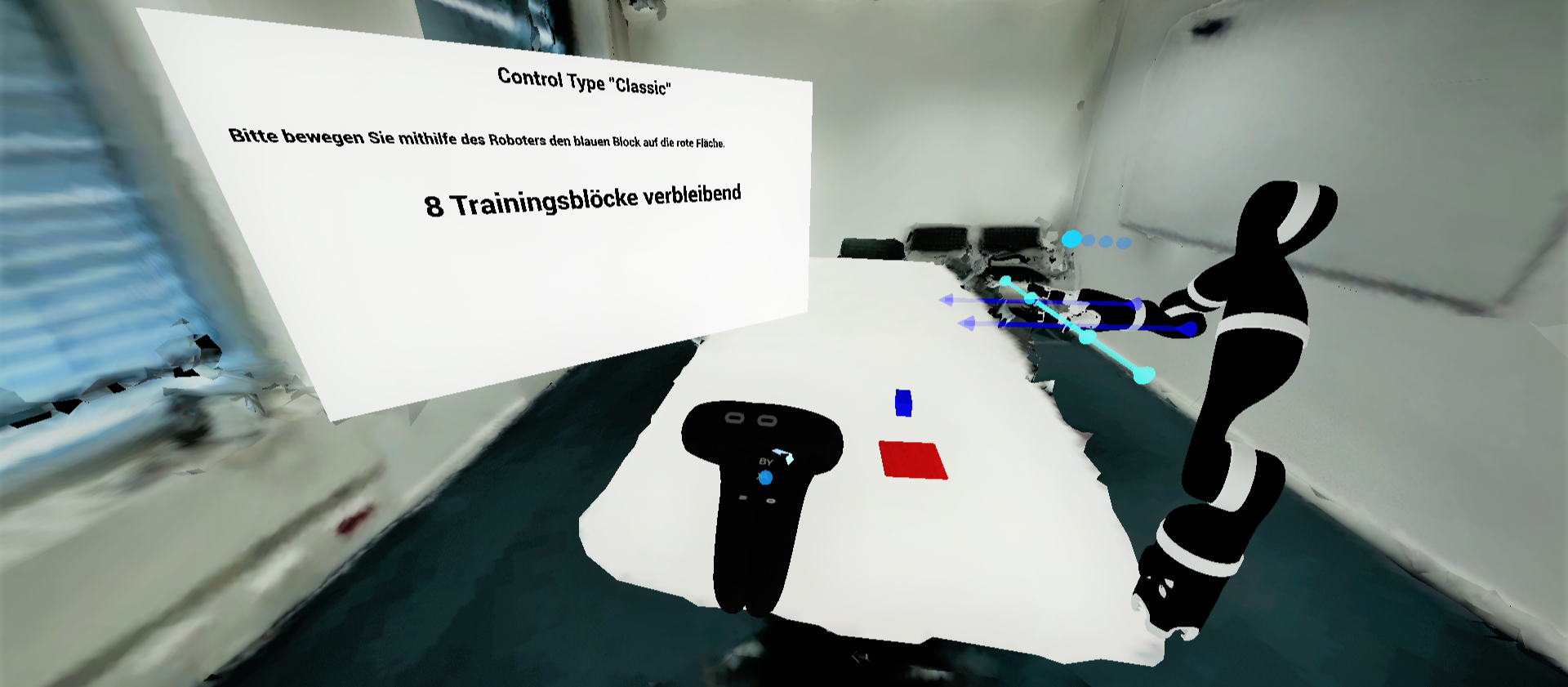}
\centering
\caption{Virtual environment consisting of (left to right): a virtual canvas, the motion controller, a table with a blue object and red target, and a \emph{Kinova Jaco} with an arrow-based visualization~\cite{Pascher2023c}.}
\label{fig:vr-environment}
\end{figure}

In the default scenario within \emph{AdaptiX}, the focus is on a room that has been meticulously scanned using photogrammetry techniques. This room contains a table with an integrated virtual robotic arm, as depicted in \autoref{fig:vr-environment}. The simulation of the robotic arm has been optimized for operation via a \ac{VR} motion controller, which features an analog stick, several functional buttons, and motion capture capabilities. An example of a compatible motion controller is the \emph{Meta Quest 2}~\cite{metaquest2}.

As most real-world scenarios will include pick-and-place operations~\cite{Fattal.2019,wang2019towards,Kim.2012,shafti2019gaze}, we designed a straightforward testbed scenario which requires to move a blue block to a red target area. 

\begin{figure}[htbp]
\centering
    \subfloat[]{\includegraphics[width=0.32\linewidth]{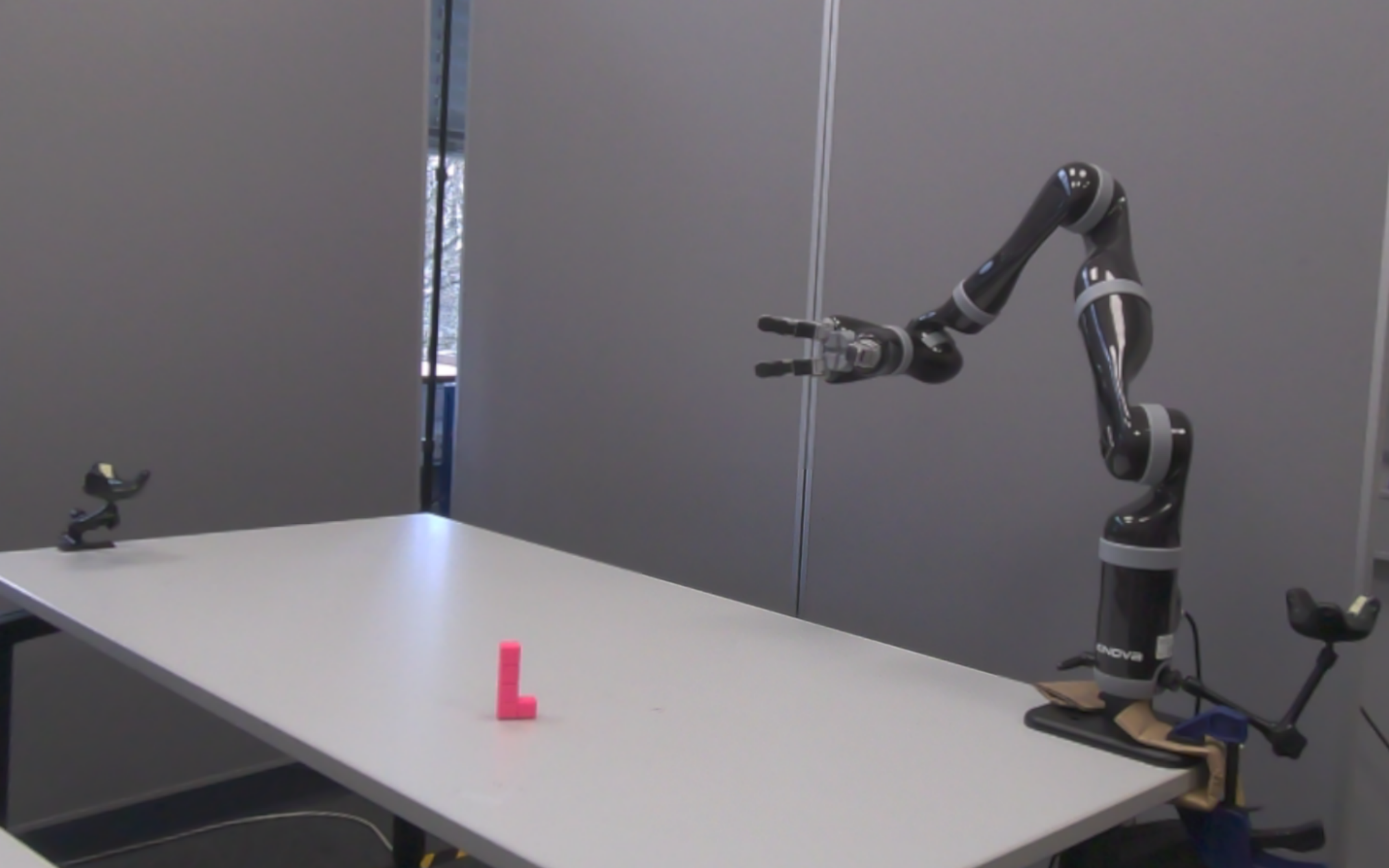}
    \label{fig:mr-continuum_a}}
    \hfill
    \subfloat[]{\includegraphics[width=0.32\linewidth]{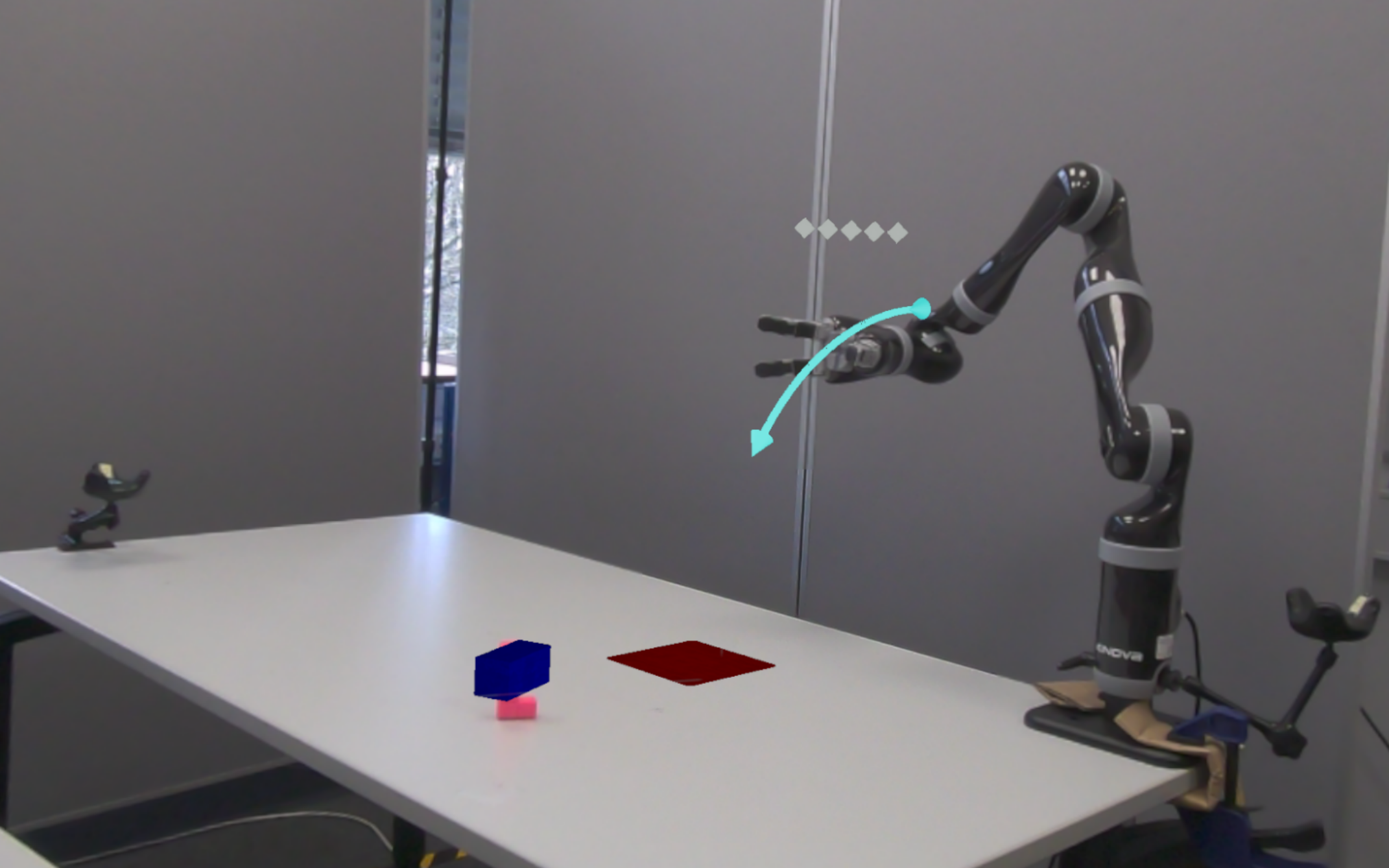}
    \label{fig:mr-continuum_b}}
    \hfill
    \subfloat[]{\includegraphics[width=0.32\linewidth]{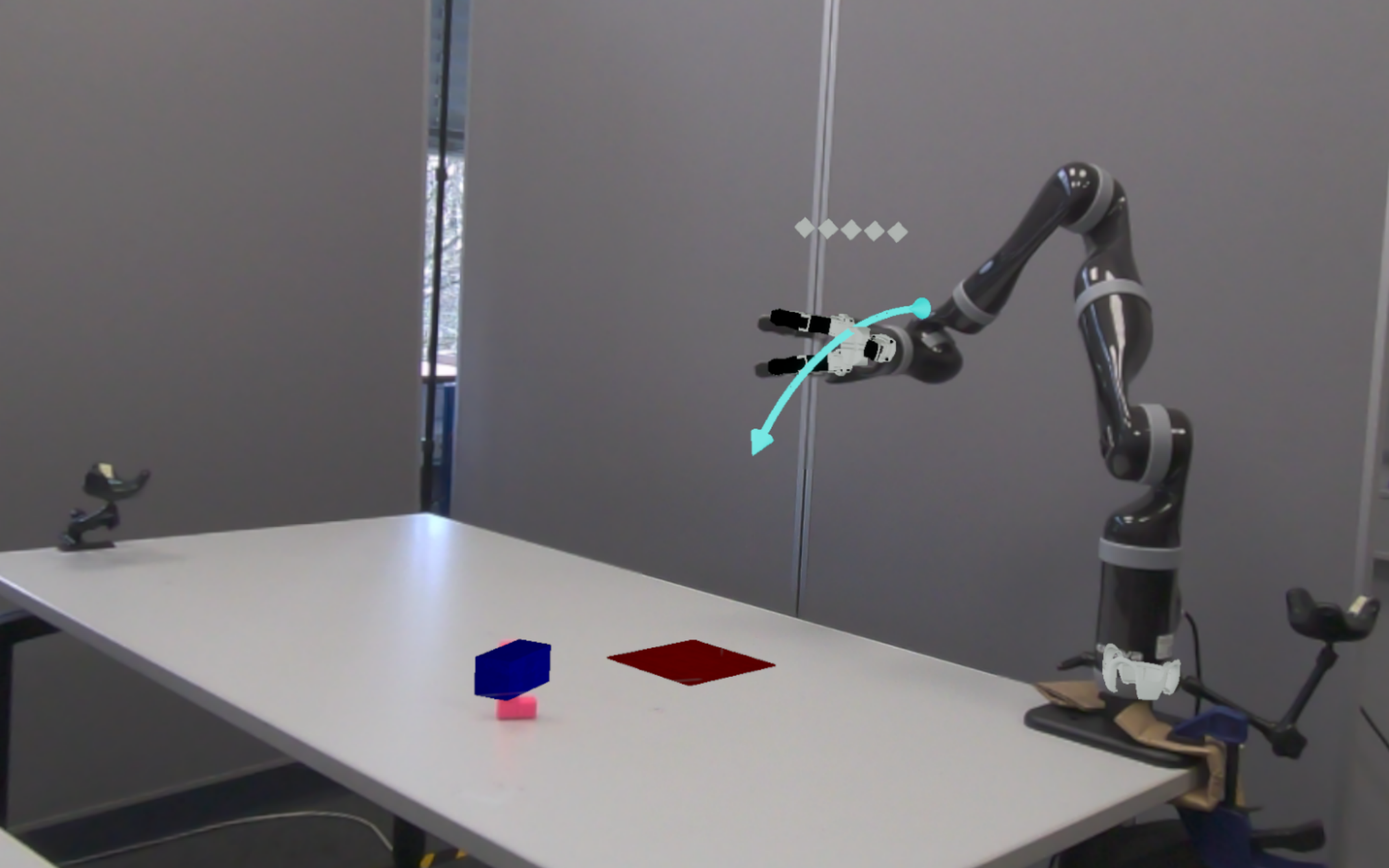}
    \label{fig:mr-continuum_c}}
    \hfill
    \subfloat[]{\includegraphics[width=0.32\linewidth]{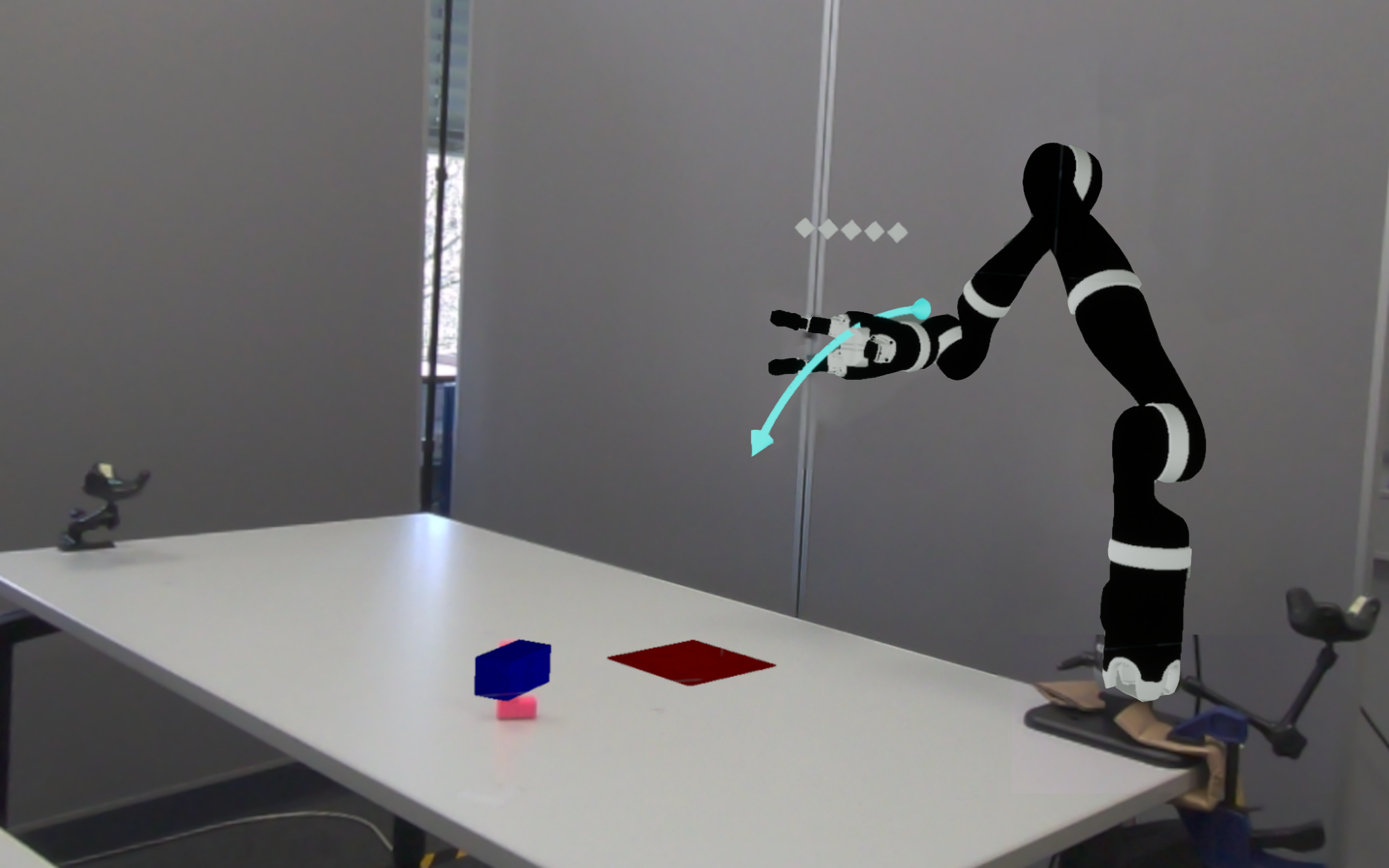}
    \label{fig:mr-continuum_d}}
    \hfill
    \subfloat[]{\includegraphics[width=0.32\linewidth]{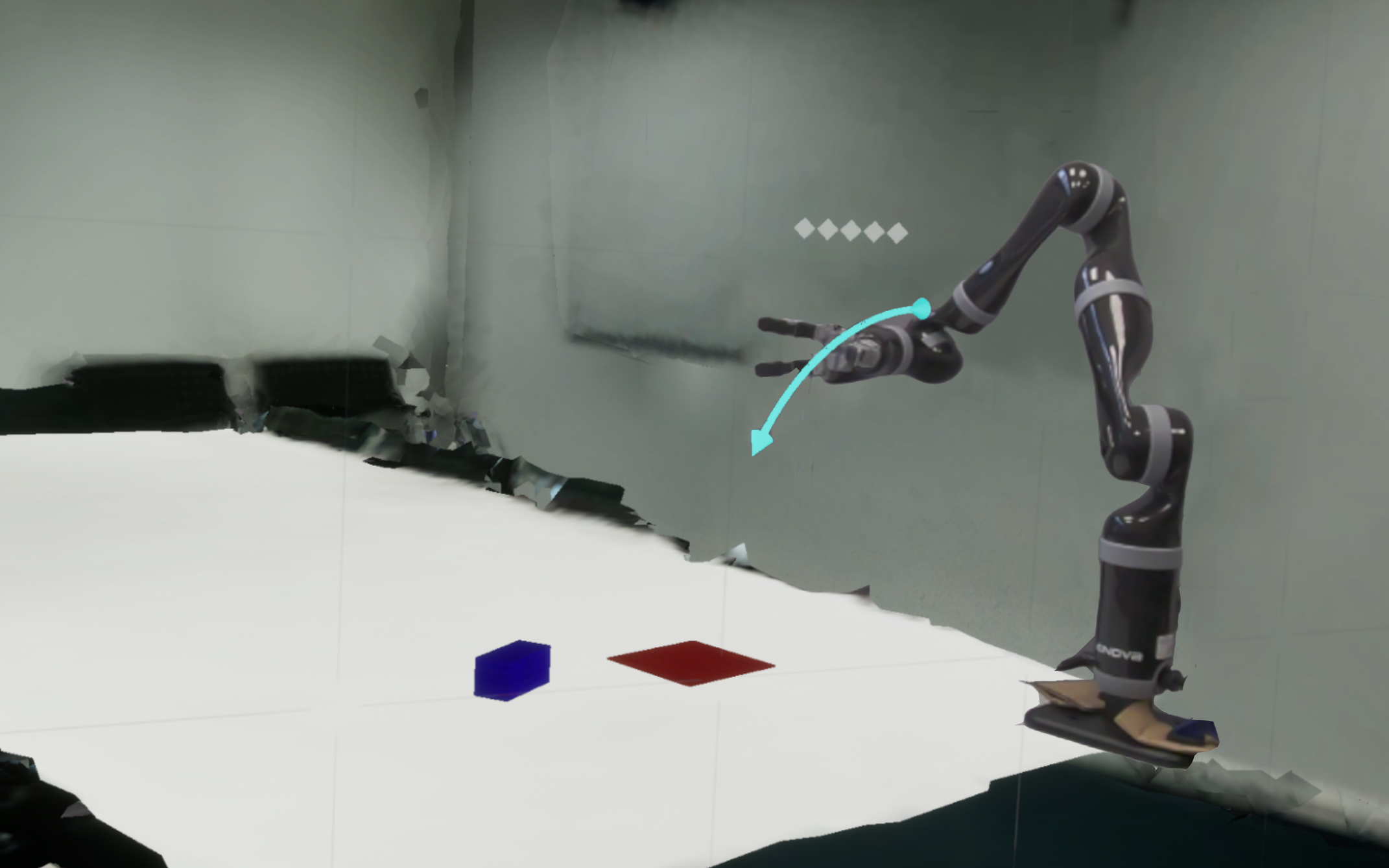}
    \label{fig:mr-continuum_e}}
    \hfill
    \subfloat[]{\includegraphics[width=0.32\linewidth]{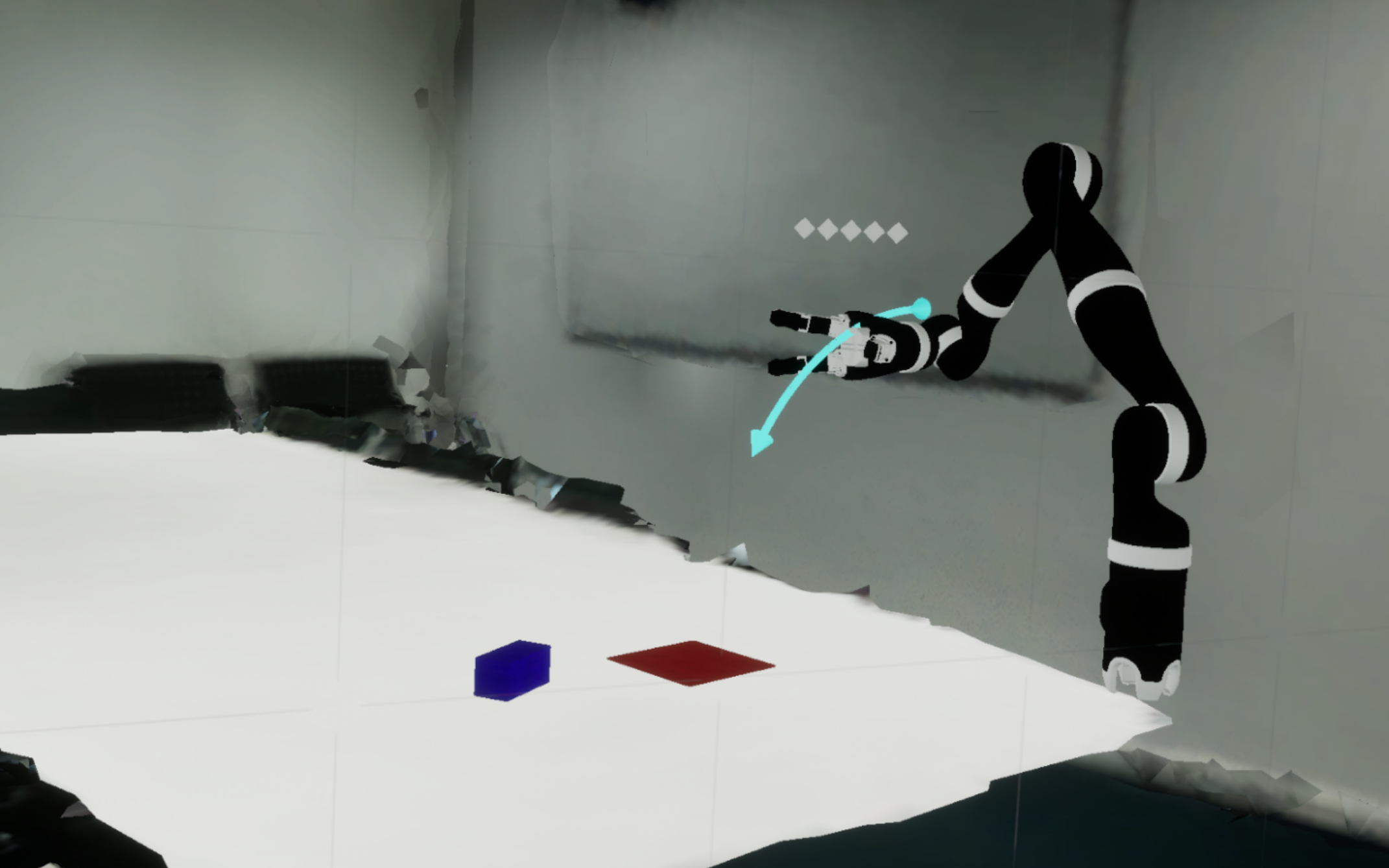}
    \label{fig:mr-continuum_f}}
\caption{\ac{MR} continuum with (\textbf{a}) only the real robotic arm in real environment, (\textbf{b}) augmenting of directional cues in the real environment with the real robotic arm, (\textbf{c}) additional visualizing the gripper and base of the virtual robotic arm in the real environment, (\textbf{d}) visualizing the simulated robotic arm in the real environment, (\textbf{e}) visualizing the real robotic arm in the virtual environment, and (\textbf{f}) visualizing the simulated robotic arm in the virtual environment~\cite{Pascher.2024adaptix}.}
\label{fig:visualizations-continuum-mr}
\end{figure}

We integrated the \emph{Varjo XR-3}~\cite{varjo}, a high-resolution \acs{XR} \ac{HMD}, to create a seamless \ac{MR} environment. By employing two \emph{HTC VIVE} trackers~\cite{vivetracker}, we synchronized the virtual and real worlds, ensuring that the operational spaces of the robots were perfectly aligned. This synchronization enables the adjustment of the MR level in multiple increments, as outlined in the \emph{virtuality continuum} proposed by Milgram and Kishino~\cite{milgram1994taxonomy}.
A visual comparison between the user's perspective in the real world and the simulation is presented in \autoref{fig:visualizations-continuum-mr}. 

The \ac{MR} continuum comprises different levels. 
Level \textbf{(1)} serves as the study's baseline condition, offering no multi-modal feedback to the user. At level \textbf{(2)}, the system mimics an \ac{AR} visualization technique, which overlays the entire physical setup with basic cues. Especially level \textbf{(3)} and \textbf{(4)} enable customizing either the robot itself or the environment to extent/exchange the physical setup but still not loosing the context. In \textbf{(3)} users can interact with a totally new or customized robot while being in a familiar environment. World's distractions can be excluded in \textbf{(4)} while the the original robot is presented. Level \textbf{(5)} provides a \ac{VR} environment that is entirely customizable.

\section{AI-enhanced Shared Control}
This section will provide more details about our previous research and how we addressed the three main challenges stated at the end of Section~\ref{sec:intro} -- \emph{\ac{AI} legibility}, \emph{\ac{AI} user control}, and \emph{\ac{AI} intervention}.
Therefore, we illustrate initial design concepts, work in progress, and prototypes that were evaluated in user studies. 

\subsection{\ac{AI} Legibility}
In the context of our research, achieving a level of \ac{AI} legibility is mostly concerned with making it easier to understand how the \ac{AI} would reassign the input mapping and/or change the movement direction of the robot. In our recent survey on such robot motion intent approaches~\cite{Pascher.2023robotMotionIntent}, we found that, for communicating location information (such as a movement direction), head-mounted technology such as \ac{AR} \acp{HMD} allow to represent the robot movement visually and have shown to provide a potential fruitful approach~\cite{cleaver_2021}. 
Although research has explored robot motion intent, there needs to be more insight into what works best in various situations and for different user types. Customizing the visualization and feedback modality is crucial, as there is no \enquote{one size fits all} solution~\cite{Holloway.2019onesizefitsone}

We proposed different design concepts that fall into a spectrum with two extremes -- indicative and explanatory~\cite{pascher.2022dof}. 
\textbf{Indicative:} Focus on crucial information only, quick and easy solution, suitable for experienced robot users. \textbf{Explanatory:} Movements are shown in great detail, high level of information, especially helpful for new users.

\begin{figure}[htbp]
    \centering
    \subfloat[]{\includegraphics[width=0.35\linewidth]{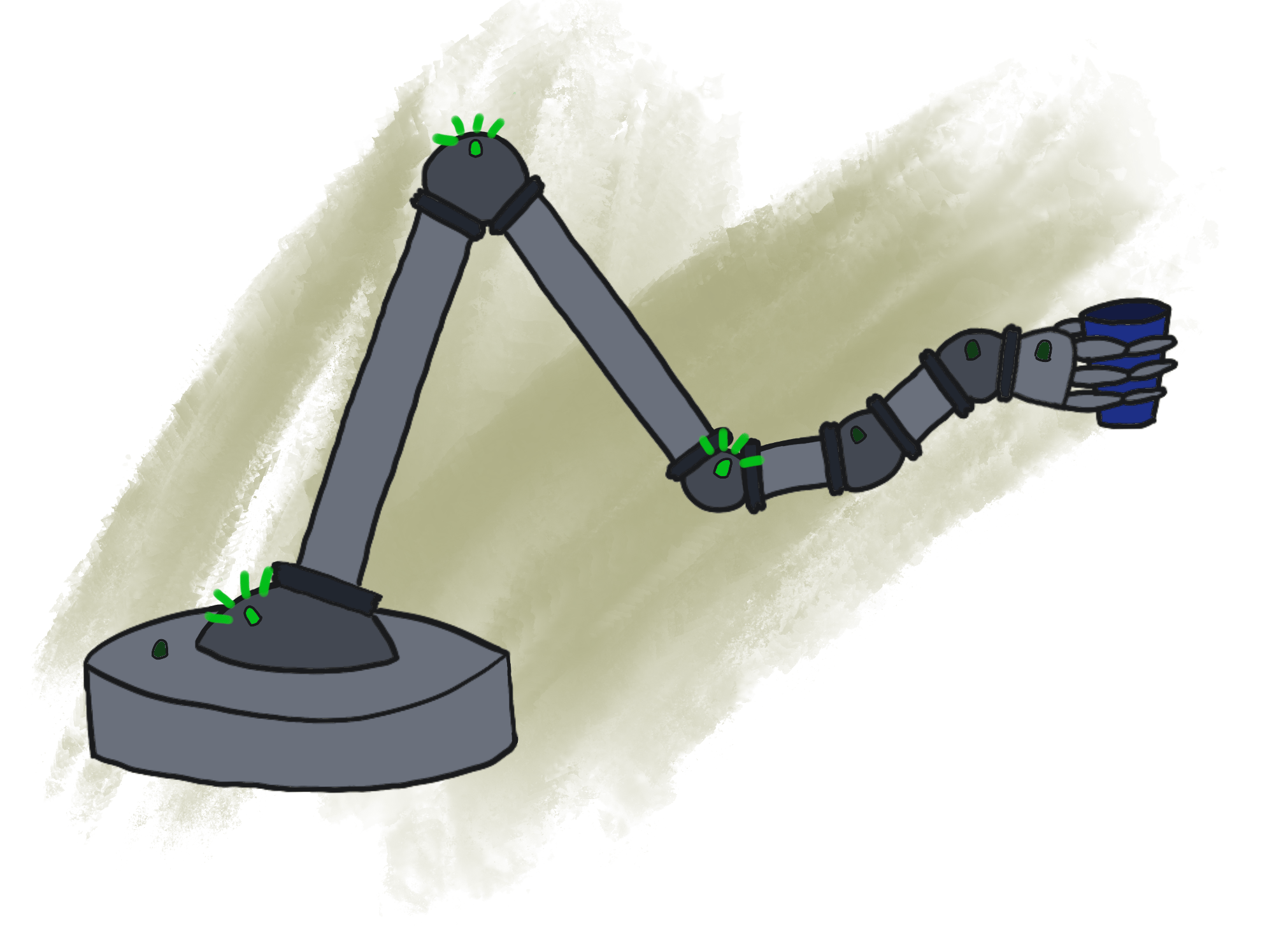}}
    \hfill
    \subfloat[]{\includegraphics[width=0.35\linewidth]{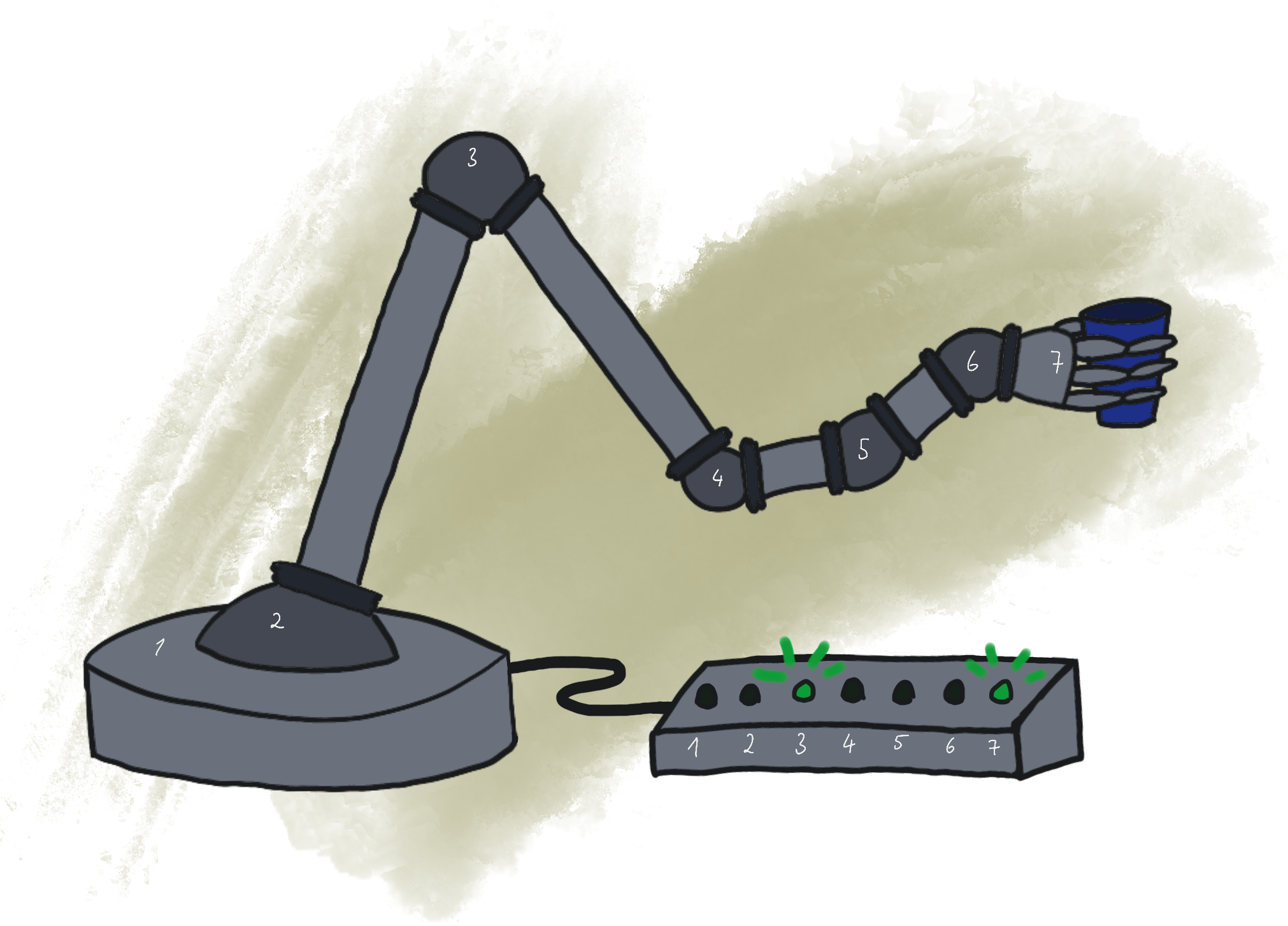}}
    \caption{DoF-Indicator: \textbf{(a)} LEDs directly attached at each robot's joint; \textbf{(b)} LEDs mounted on a bar in front of the robot referring to each joint (1--7)~\cite{pascher.2022dof}.}
    \label{fig:led}
\end{figure}

\textbf{DoF-Indicator:} LEDs attached to the robot's axis and joints -- or mounted on a bar in front of it -- communicate active and nonactive \acp{DoF} (see~\autoref{fig:led}). Likely more suitable for experienced users because of indirect communication of movement, it communicates the \ac{DoF} mapping and resulting movement abilities.

\begin{figure}[htbp]
    \centering
    \subfloat[]{\includegraphics[width=0.35\linewidth]{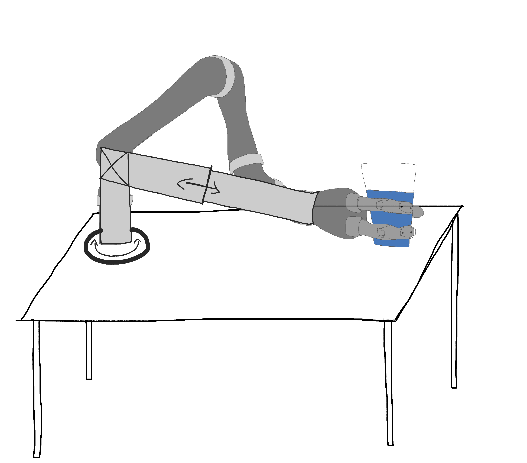}}
    \hfill
    \subfloat[]{\includegraphics[width=0.35\linewidth]{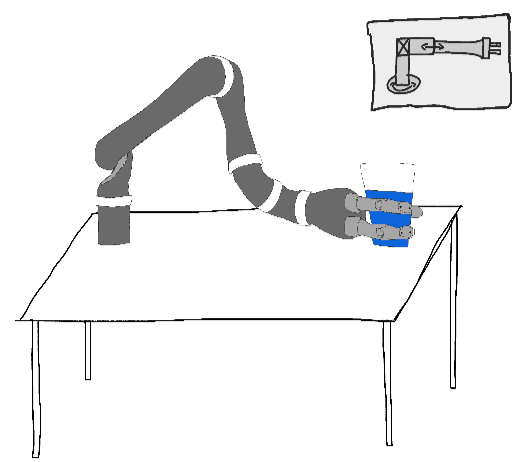}}
    \caption{DoF-Combination-Indicator: \textbf{(a)} as an \ac{AR} overlay, supporting robot and visualization in line of sight; \textbf{(b)} as an icon in the screen's corner~\cite{pascher.2022dof}.}
    \label{fig:fakejoint}
\end{figure}

\textbf{DoF-Combination-Indicator:} Movement ability is communicated by a simplified representation of the robot only able to move two \acp{DoF} simultaneously, for example, rotating and extending the arm (see~\autoref{fig:fakejoint}). The \ac{AR} representation either overlays the real robot or can be displayed separately in the corner of the \ac{AR} screen. This visualization decreases the robot's complexity.

\begin{figure}[htbp]
    \centering
    \includegraphics[width=\linewidth]{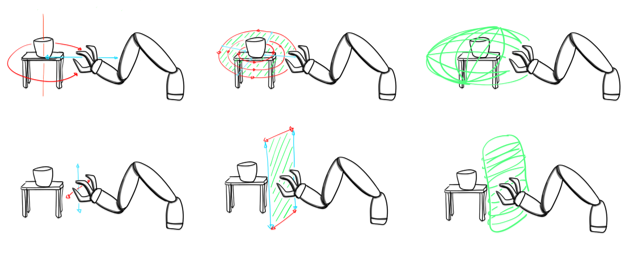}
    \caption{Gizmo Visualization: \textbf{(left)} simple: straight and curved arrows; \textbf{(center)} planar: planes of movement; \textbf{(right)} cloud: 3D-cloud of possible positions~\cite{pascher.2022dof}.}
    \label{fig:gizmo}
\end{figure}

\textbf{Gizmo Visualization:} Arrows, planes and point clouds communicate the current movement ability of the robot (see~\autoref{fig:gizmo}). This allows for several different design options. A  arrow-based approach was already successfully evaluated in previous studies~\cite{Kronhardt.2022adaptOrPerish,Pascher2023c}.

\textbf{Demonstration:} Current movement possibilities are demonstrated through either the actual robot or an \ac{AR} ghost-representation. With both options a quick movement indicates the intended motion. 

\subsubsection{Visualization Approaches}
\begin{figure}[tb]
\centering
  \subfloat[]{\includegraphics[width=0.43\linewidth]{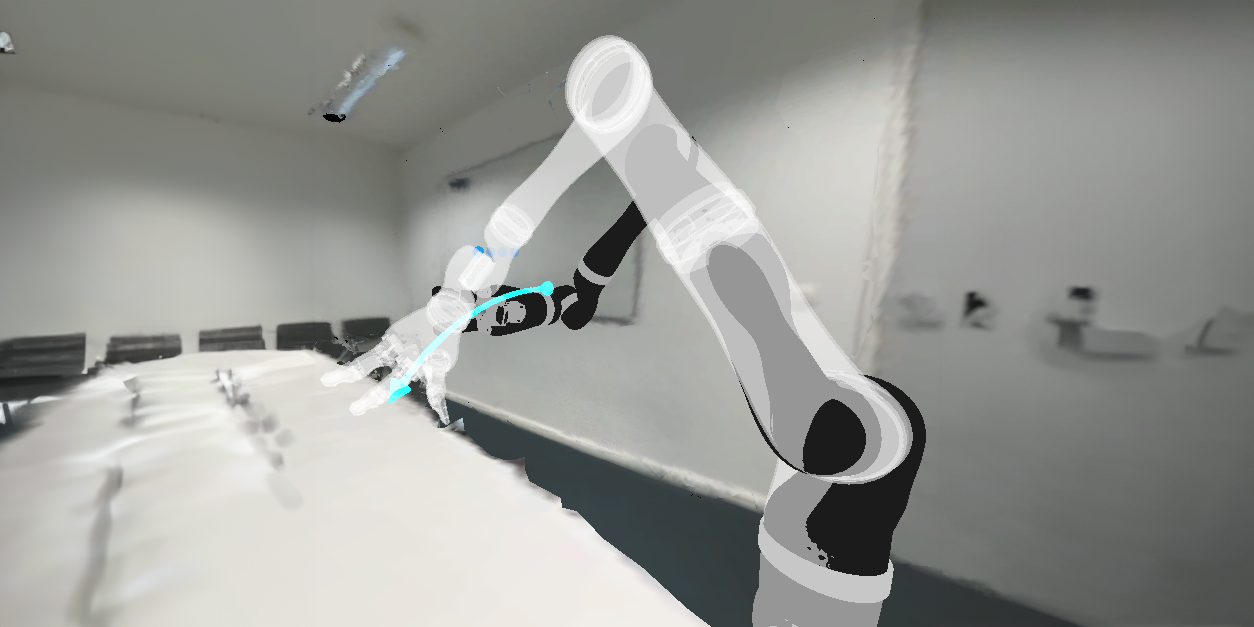}}
  \hfill
  \subfloat[]{\includegraphics[width=0.43\linewidth]{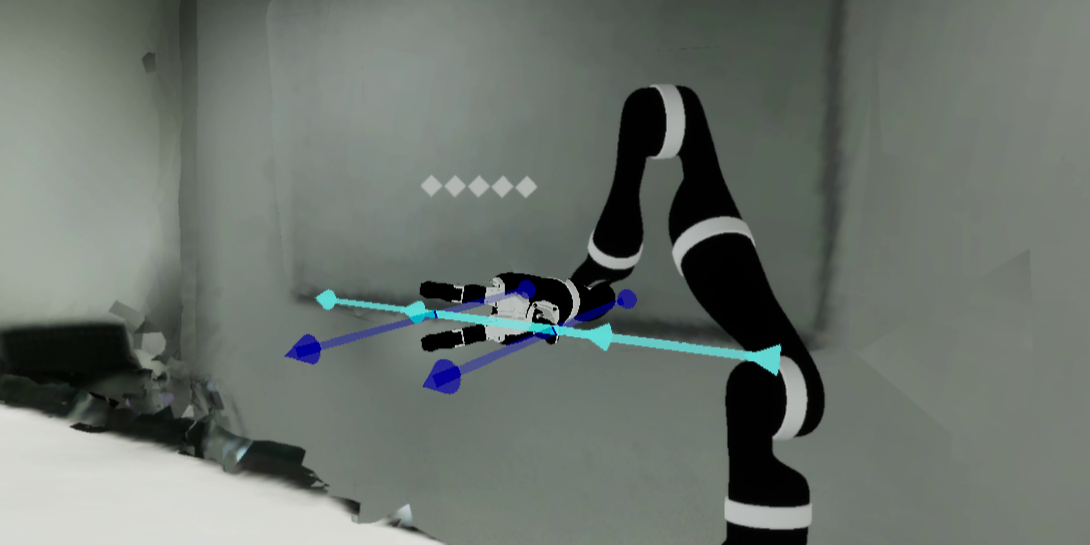}}
  \hfill
  \subfloat[]{\includegraphics[width=0.43\linewidth]{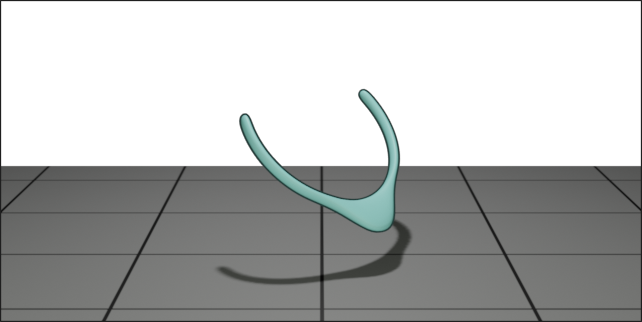}}
  \hfill
  \subfloat[]{\includegraphics[width=0.43\linewidth]{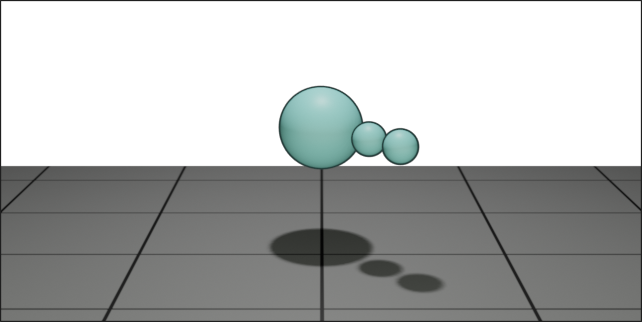}}
\centering
\caption{Visualization examples for directional cues of an \ac{AI}-supported robotic control in a \ac{VR} 3D environment: (\textbf{a}) Ghost \& (\textbf{b}) Arrows~\cite{Pascher.2024adaptix}, and a real-world setting via data glasses, e.g., \emph{Google Glass EE2}: (\textbf{c}) Ring \& (\textbf{d}) Points.}
\label{fig:visualizations-interfaces}
\end{figure}

Based on our initial concepts, we have explored augmenting the users' view with directional movement cues both in true -- three-dimensional -- \ac{AR} (registered in 3D,~\cite{Kronhardt.2022adaptOrPerish,Pascher2023c}) as well as in 2D as symbolic representations on a data glass (for both refer to \autoref{fig:visualizations-interfaces}). 

The former allows information-rich visualization and has shown in our studies to allow users to sufficiently anticipate a new suggested input mode mapping and corresponding movement direction~\cite{Pascher2023c}. 

The latter provides the advantage that the technology is already market ready and the devices are lightweight to carry and relatively easy to set up (compared to \ac{AR} \acp{HMD}). However, without the ability to display directional visual cues registered in 3D space, the visual feedback is separated from the interaction space (robot) and may be more difficult to align with the current robot movement. In our work, we are currently exploring different visual forms, as can be seen in \autoref{fig:visualizations-interfaces} c) and d).

\begin{figure}[htbp]
\centering
\captionsetup{justification=centering}
\hfill
    \subfloat[3D Gradient]{\includegraphics[width=0.32\linewidth]{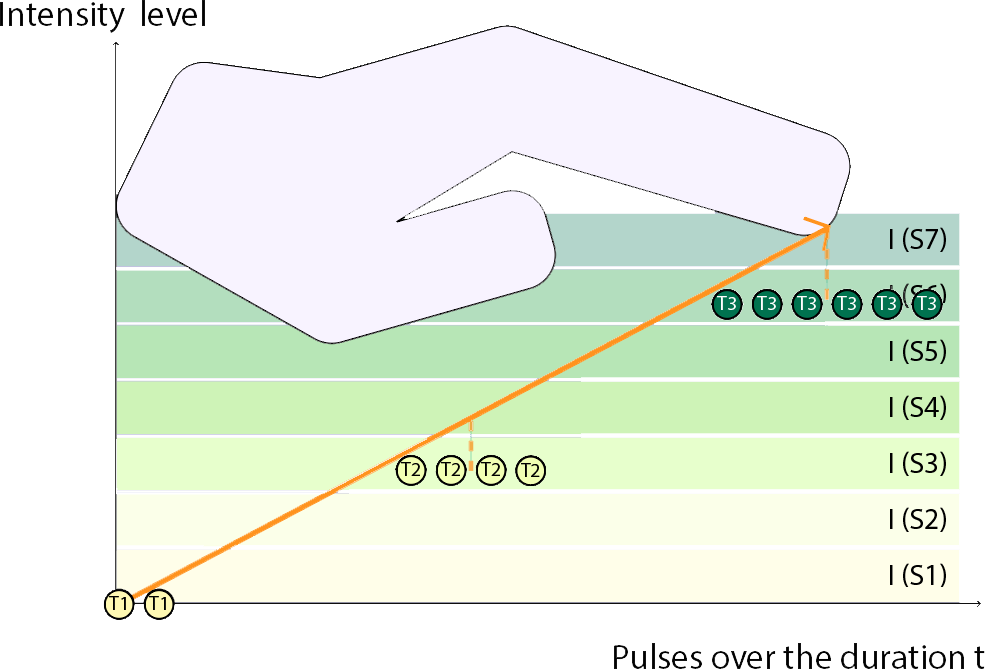}\label{fig:rabbit}}
    \hfill
    \subfloat[2D Directions]{\includegraphics[width=0.28\linewidth]{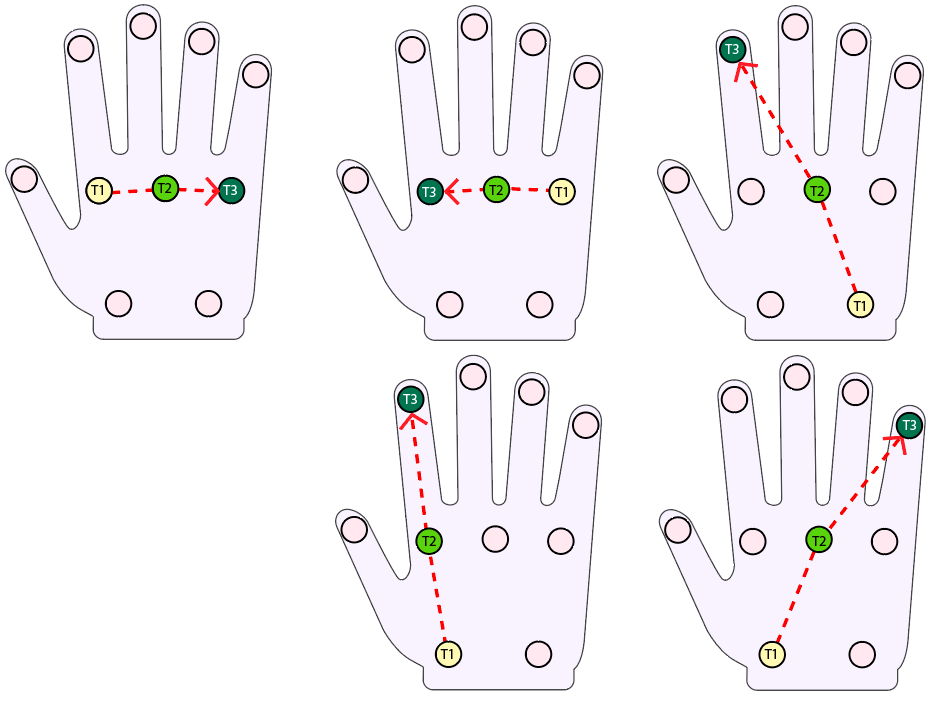}\label{fig:directions}}
    \hfill
    \subfloat[Study Setup]{\includegraphics[width=0.32\linewidth]{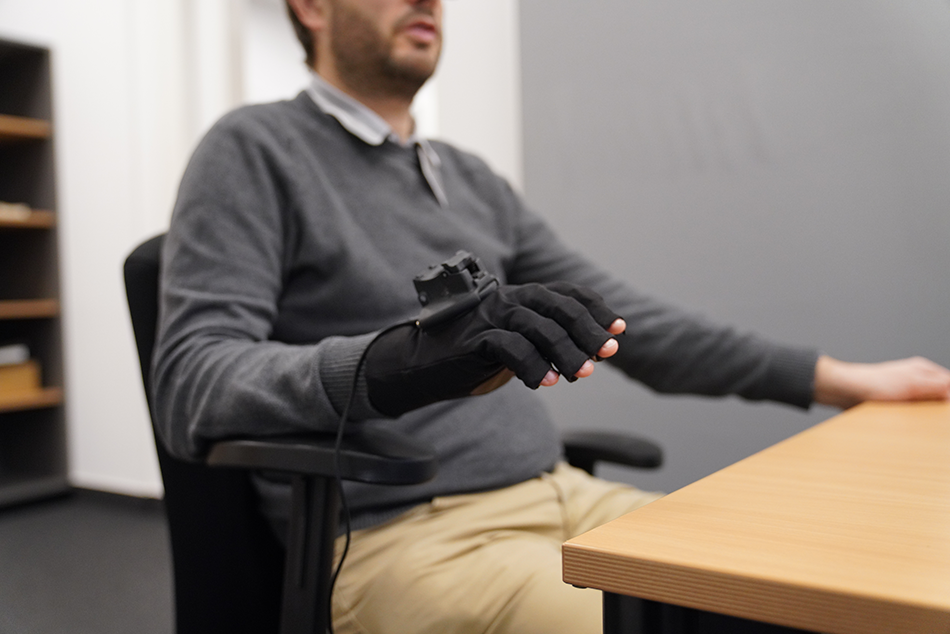}\label{fig:setup}}
    \hfill
\captionsetup{justification=justified}
  \caption{Vibrotactile directional cues \textbf{(a)} 3D Gradient encoding with pulses and intensity mapping \textbf{(b)} 2D direction encoding across the hand \textbf{(c)} illustrates the study setup, with the arm resting on the armrest while the hand is in the air~\cite{Pascher2023d}.}
  \label{fig:hand-setting}
\end{figure}

In addition, we have been exploring other modalities which could either replace or complement visual feedback to increase the legibility of the \ac{AI}. In~\cite{Pascher2023d}, we explored different designs for vibrotactile feedback to communicate three-dimensional motion directions. We developed two conditions based on the \emph{Cutaneous Rabbit} illusion and one based on \emph{Apparent Tactile Motion} to communicate 2D direction. The gradient of the overall 3D direction was then encoded by the number of discrete vibration pulses, the vibration intensity, or a combination of both. Our study showed that three-dimensional directional cues could be communicated with a high success rate for both the 2D direction and gradient, but may benefit from dual-encoding of the gradient information as well as individual customization of the specific implementation of the vibrotactile feedback patterns (see \autoref{fig:hand-setting}).

\subsubsection{Empirical Implications}
Through related work and our own research, we identified that -- while the robot system is designed to act in the user's best interest -- the user still needs to build trust, which requires transparency and legibility that they can comprehend. They should also be able to interfere with the robot's control if the robot makes a mistake or gives inappropriate suggestions for interaction. Communicating intent further requires having the user pay attention or guiding the attention of the user, requiring multi-modal stimuli, depending on the situation and the capabilities of the user.

\subsection{\ac{AI} User Control}
For the user to remain in control, automatic mapping of input modes may not be desirable. Instead, we have explored different ways which allow users to stay in control but also benefit from the potential increase in efficient task completion. In their original approach, Goldau \& Frese~\cite{Goldau.2021petra} asked users to wait for five seconds without interacting with the robot to trigger a new mapping. 

As this has shown to cause some level of frustration, we explored ways for the user to directly request a new mapping~\cite{Kronhardt.2022adaptOrPerish} as well as integrating a continuous or threshold-based feedforward visualization of an updated mapping~\cite{Pascher2023c}. 
The latter has shown positive effects, as the user thereby can always compare the current movement and mapping with an update from the \ac{AI}, but only switch to that when they feel that it changes the movement direction to achieve the task better.

In both studies, \emph{Classic} -- a non-adaptive control mode inspired by the \emph{Kinova Jaco 2} standard joystick input -- relies on mode switching to access and control all \acp{DoF} one after another and was used as a baseline condition. 
In comparison to \emph{Classic}, our \ac{AI}-based \ac{ADMC} methods significantly reduced \textbf{(1)} the task completion time, \textbf{(2)} the average number of necessary mode switches, and \textbf{(3)} the perceived workload of the user.

Users may have diverse input device preferences and capabilities. This calls for the availability of multi-modal input options or the ability to choose between different input modalities~\cite{Arevalo-Arboleda2021b}.
To enable such user input, our simulation environment provides a standard control approach where pressing a keyboard button moves the end effector along cardinal \acp{DoF} (x, y, z, roll, pitch, yaw, opening and closing the gripper). Using further build-in functionalities, the designated keyboard input can easily be adjusted to other input devices like gamepads, joysticks, or customized assistive input appliances.

\subsubsection{Empirical Implications}
While the goal is to keep users in control, the complexity of both the robot interaction and the \ac{DoF} limitations for available input devices can easily make the system very difficult to use. Therefore, in order to find the sweet spot for shared control, we propose to start with a rather minimized set of user interaction and increase that on demand and depending on the individual capabilities. Users should not be confused by too many interaction options or overly complex movements. While optimal ways of accomplishing a goal may require complex intervention from the robot, these interventions may be difficult for users to understand, and therefore trust. In addition, keep the \ac{DoF} of input devices low as this maximizes the amount of assistive devices capable of controlling the robot.

Our previous research and related work show that pick-and-place tasks are ubiquitous and necessary to perform \ac{ADLs}. It is, therefore, important that shared control is first implemented for these simple tasks before more complex sequences are examined. If users struggle to understand shared controls for pick-and-place tasks, we believe it is highly likely that more complex tasks may cause further frustration.

\subsection{\ac{AI} Intervention}
While our approaches for \ac{AI} user control allow some level of intervention, since the user can decide when to accept the updated input mapping as provided by the \ac{AI}, it has some limitations. If the \ac{AI} algorithm -- as suggested by Goldau \& Frese -- is not able to provide a useful mapping, the user may become stuck with little flexibility to trigger the \ac{AI} to update the mapping. 
We are currently exploring different approaches to tackle this issue for the specific shared control mechanism. One rather straightforward approach would be to allow the user to disable the \ac{AI} and go back to manual mode switching of Cartesian \acp{DoF}. This of course may decrease the acceptance and perceived usefulness of the \ac{AI}. 

\begin{figure}[htbp]
    \centering
    \includegraphics[width=\textwidth]{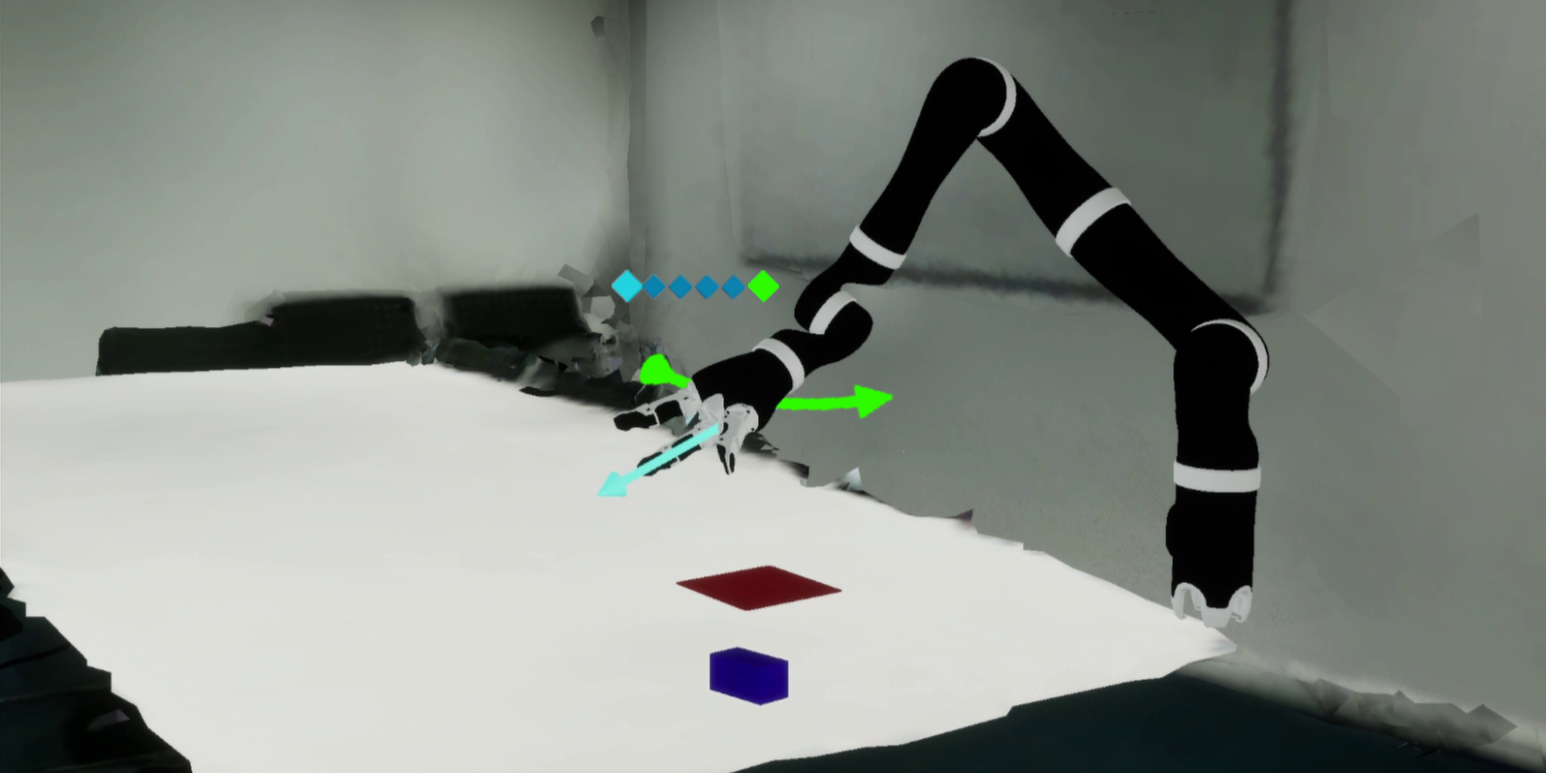}
    \caption{The current \ac{DoF} mapping (cyan arrow) does not allow to move to the blue object. By \emph{changing the perspective} (green mode and arrow), the gripper is rotated in place to allow the \ac{CNN} to suggest a new \ac{DoF} mapping by an updated camera feed.}
    \label{fig:reposition-jaco}
\end{figure}

A different way, which directly builds on the understanding of how the \ac{AI}, in this case a \ac{CNN}, operates, would be to find a way for the \ac{AI} to \emph{change perspective} -- quite literally (see \autoref{fig:reposition-jaco}). If the robotic arm, or more specifically the gripper with the integrated or mounted camera is triggered to perform a small location repositioning, basically resulting in the robot looking around, the \ac{CNN} will receive a new input which may result in a new and potentially better mapping.

Especially when the \ac{CNN} is confronted with more than one choice (e.g., several objects in sight and vicinity of the gripper/camera) and has to choose one of them, based on the \ac{AI} algorithm, rather than resulting in a draw. This kind of deadlock would only be resolved by a manual -- Cartesian \ac{DoF} -- control. More suitable would be to either increasing \ac{AI}'s confidence of the chosen object by user input in the selected direction or decreasing by moving away, leading to a re-calculation and updated \ac{DoF} mapping suggestion.

\subsubsection{Empirical Implications}
The aim is to keep the user in the loop so that they can intervene appropriately whenever the \ac{AI} reaches its limits. However, this must strive for a balance that does not place sole decision dependency on the user to avoid access cognitive demand and temporal delays. Instead, establishing a four-eye principle with the \ac{AI} functioning with implicit user consent until intervention is the most efficient approach to fulfilling the task's goal.

\section{Conclusion}
In this paper, we summarized our experiences for engineering \ac{AI}-enhanced shared-control methods for assistive robotic arms. In particular, we identified three main challenges in \emph{\ac{AI} legibility}, \emph{\ac{AI} user control}, and \emph{\ac{AI} intervention}. Our work highlights the benefits and importance of sensible interaction design, which addresses these challenges and requires both a deep understanding of and interconnection with the \ac{AI} technology. We also found that there is still much to be explored, in particular in the area of \emph{\ac{AI} intervention} approaches which go beyond circumventing the \ac{AI}.

\bibliographystyle{splncs04}
\bibliography{MS}
\end{document}